\documentclass[aps,prl,twocolumn,superscriptaddress,floatfix,longbibliography]{revtex4-1}

\usepackage{amssymb}
\usepackage{graphicx}
\usepackage{dcolumn}
\usepackage{bm}
\usepackage{amsmath}
\usepackage{ulem}
\usepackage[colorlinks,linkcolor=magenta,citecolor=blue,urlcolor=blue]{hyperref}

\begin{document}

\title{Fate of Two-Particle Bound States in the Continuum in Non-Hermitian Systems}
\author{Yanxia Liu}
\affiliation{School of Physics and Astronomy, Yunnan Key Laboratory for Quantum Information, Yunnan University, Kunming 650091, People's Republic of China}

\author{Shu Chen}
\email{schen@iphy.ac.cn}
\affiliation{Beijing National Laboratory for Condensed Matter Physics, Institute of
Physics, Chinese Academy of Sciences, Beijing 100190, China}
\affiliation{School of Physical Sciences, University of Chinese Academy of Sciences,
Beijing, 100049, China}

\begin{abstract}

We unveil the existence of two-particle bound state in the continuum (BIC) in a one-dimensional interacting nonreciprocal lattice with a generalized boundary condition. By applying the Bethe-ansatz method, we can exactly solve the wave function and eigenvalue of the bound state in the continuum band, which enable us to precisely determine the phase diagrams of  BIC. Our results demonstrate that the nonreciprocal hopping can delocalize the bound state and thus shrink the region of BIC. By analyzing the wave function, we identify the existence of two types of BICs with different spatial distributions and analytically derive the corresponding threshold values for the breakdown of BICs. The BIC with similar properties is also found to exist in another system with an impurity potential.


 \end{abstract}

\maketitle


{\it Introduction-} An intriguing property of non-Hermitian systems is the sensitivity of wave functions and spectra to boundary conditions and local impurity. Intensive studies have unveiled that non-Hermitian nonreciprocal systems can exhibit some novel non-Hermitian phenomena, e.g., non-Hermitian skin effect (NHSE) \cite{LeeCH,13,15,16,17,18} and scale-free localization \cite{LiLH2021,Murakami,GuoCX2023,WangZ2023,Bergholtz2023}, characterized by the emergence of diverse localizing behaviors and changes of spectrum structures when translational invariance is locally broken, either by introducing an impurity or by tuning the boundary coupling strength \cite{LiuYX2021,CXGuo2021}. While most novel phenomena and concepts about the non-Hermitian effects are built based on the non-interacting systems, recently
there is growing interest in exploring non-Hermitian phenomena in interacting systems \cite{XuZ,ZhangDW,Nori2020,Fukui,ZhengMC,LiLH2024}.

The concept of bound state in the continuum (BIC) was initially proposed by von Neumann and Wigner, which refers to the eigenenergies of the bound states that can be embedded in the continuum \cite{Neumann29,Hsu16}. Recently,  BICs attracted much renewed interest both theoretically and experimentally \cite{Zhang23,Zhen14,Kang23,Qian24,Vaidya21,Cerjan20,Yuan20,Xiao17}. The BICs are found to be present in many physical systems, including the Hubbard model\cite{ZhangJM12,ZhangJM13,Valle14,YAshida22,Sugimoto23,WZhang23,Huang23}, optical systems \cite{Koshelev18,Cao20,Kravtsov20,Fan19,Han19}, etc, and have caused many applications such as enhanced optical nonlinearity \cite{Carletti18,Bulgakov14,Krasikov18}, sensing \cite{Yanik11,Liu17,Romano19}, lasers \cite{ Kodigala17,Song18,Midya18,Yu21}, and filtering \cite{Foley14}.
 Usually, the emergence of BIC in a single particle system requires certain exotic potentials. For multiparticle systems, BIC can be created via the interplay of local potential and particle-particle interaction \cite{ZhangJM12,ZhangJM13,Sugimoto23,WZhang23}.
It has been demonstrated that BICs can be engineered in impurity systems by adding interparticle interaction \cite{ZhangJM12,ZhangJM13}. Meanwhile, edge states in the continuum are also found in interacting topological models \cite{Liberto16,Gorlach17}. Usually, bound states must have quantized energies in a Hermitian system, whereas free states form a
continuum.  However, this principle may fail for non-Hermitian systems \cite{MaGC}. Some recent works demonstrate that continuum of bound states can occur in non-Hermitian systems \cite{MaGC,ChongYD,ChenG}.
In the presence of NHSE, a local bound state may be delocalized depending on the competition between nonreciprocal hopping and impurity strength. Studies on the impurity model in nonreciprocal lattices reveals that the bound state disappears when it touches the continuum of spectrum \cite{LiuYX2020}. As most previous studies of BIC focused on the Hermitian systems, this raises our interest to pursue the BIC in non-Hermitian system with nonreciprocal hopping and interaction.

In this Letter, we study two-particle BICs in non-Hermitian lattices and unveil their specific feature and fate under the influence of the nonreciprocal hopping. To be concrete, we consider two interacting bosons in a nonreciprocal lattice with GBC or an impurity potential. Although this model is nonintegrable, we demonstrate that the BIC can be obtained analytically with a similar Bethe-ansatz method.
The analytical expression of wave function and bound energy enables us to determine the phase diagram of BIC exactly. We unveil the existence of two kinds of BICs, characterized by different spatial distributions, which are delocalized by increasing the nonreciprocal hopping. By analyzing the wave function, we derive analytical expressions of threshold values for the breakdown of BICs. Our analytical results provide a firm ground for understanding the fate of BICs in non-Hermitian systems.

\begin{figure}[tbp]
\includegraphics[width=0.48\textwidth]{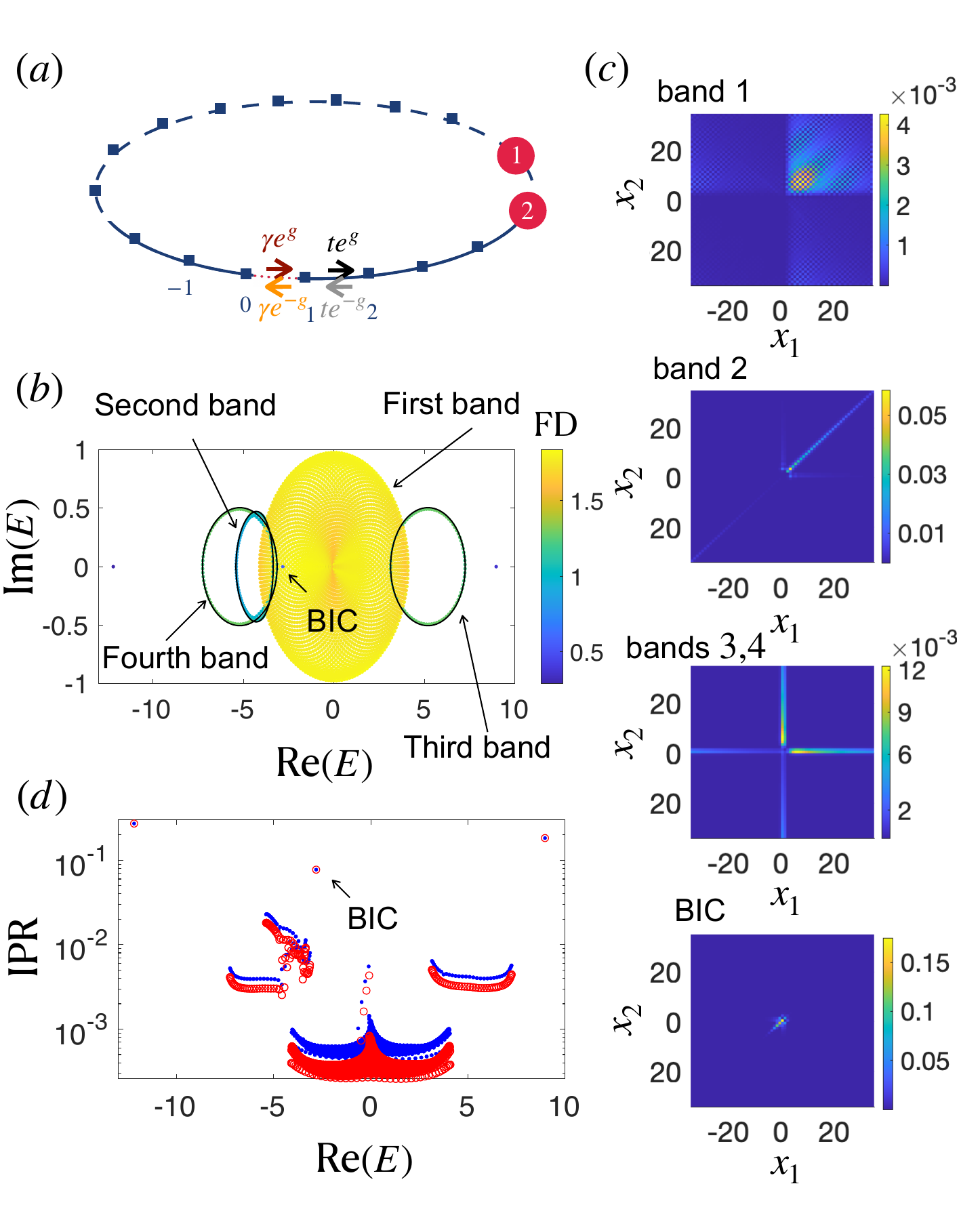}
\caption{(a) Schematics of the two-particle Hatano-Nelson model with GBCs described in Eq. \eqref{Hamiltonian}. The dark blue squares represent the lattice. The red circles represent two particles. (b)  The real and imaginary parts of spectra of the system with parameters $(g,\gamma,U)=(0.25,5,-3.5)$. The color of each dot represents the fractal dimension FD of different eigenstates. The three black circles from left to right are the fourth band  $E_{4}=-(\gamma+1/\gamma)+2\cos (k-i g)$,  the second band $E_{2}=-\sqrt{U^2+16\cos^2(K-ig) }$, and the third band $E_{3}=\gamma+1/\gamma+2\cos (k-i g)$, respectively. The length of the lattice is $N=2M=90$. (c) From top to bottom, we display the density distribution of the wave function in the first, second, third (or fourth) bands and BIC, respectively. (d). IPRs  for eigenstates with the corresponding real parts of eigenvalues $Re(E)$  for the system with parameters $(g,\gamma,U)=(0.25,5,-3.5)$. The length of the lattice is chosen to be $N=70$ (blue dots) and $N=90$ (red circles). }
\label{fig1}
\end{figure}


{\it Model and spectrum.-} The model we consider consists of two identical spinless bosons in the Hatano-Nelson model with generalized boundary conditions (GBCs), as illustrated in Fig \ref{fig1} (a), which can be described by
 \begin{align}\label{Hamiltonian}
 \hat{H}&=\sum_{x=-L,x\neq 0} ^{M}  -t\left(e^{g} \hat{a}_x^{\dag}  \hat{a}_{x+1} + e^{-g} \hat{a}_{x+1} ^{\dag}  \hat{a}_x\right)    \notag\\
& +\sum_{x=-L} ^{M}\frac{U}{2} \hat{a}_x^{\dag}  \hat{a}_x^{\dag}  \hat{a}_x  \hat{a}_x -\gamma \left(  e^{g} \hat{a}_0^{\dag}  \hat{a}_{1} + e^{-g} \hat{a}_{1} ^{\dag}  \hat{a}_0  \right) ,
\end{align}
 where $\hat{a}_x^{\dag}$ $( \hat{a}_{x})$ is the boson creation (annihilation) operator at site $x$, $te^g$ and $te^{-g}$ are imbalanced hopping amplitudes, $\gamma$  represents the boundary strength, and $t,~\gamma,~g \in \mathbb{R}$.  $U$ is the interaction strength between particles and we shall consider the case of attractive interaction with $U<0$. Cases of $U>0$ are symmetric to cases of $U<0$, thus we omit them for brevity.  We set $t=1$ as the unit of energy in the following calculation.
 For clear numerical presentation, we put the boundary in the middle of the lattice, and we label the lattice sites from $-L$ to $M$ and set $\hat{a}_{-L}^{\dag}=\hat{a}_{M+1}^{\dag}$, $\hat{a}_{-L}=\hat{a}_{M+1}$. $L=M-1$ for even lattice, $L=M$ for odd lattice.

\begin{figure*}[tbp]
\includegraphics[width=0.96\textwidth]{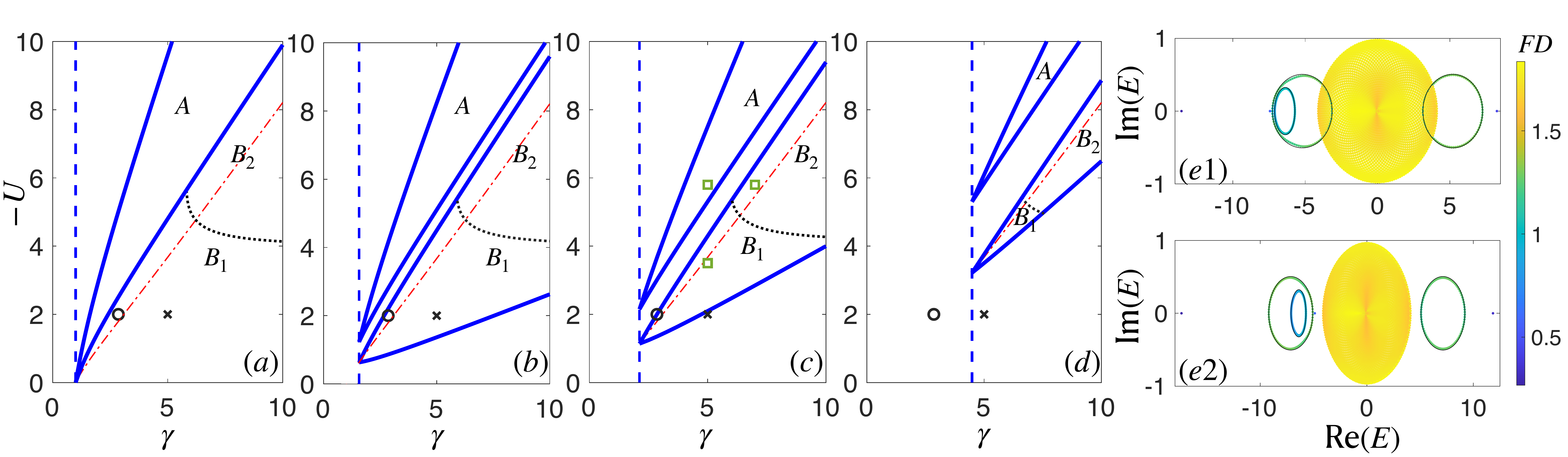}
 \caption{(a)-(d) The phase diagram for the bound state \eqref{wave functionbound} with parameter $g=0$, $0.15$, $0.25$, and $0.5$, respectively. The bound state with $E_{b1}$ exists in the region $A$, which is below all four real parts of continuum bands. In the region of $B_1$, the state with energy $E_{b2}$ is a bound state in the first continuum band $2\cos(k_1+ig)+2\cos(k_2+ig)$. In the region of $B_2$, this state is below the real part of the first and third continuum bands and above the real part of the second and fourth continuum bands.  The condition that distinguishes those two regions is $E_{b1}=4\cosh g$, which is represented by the black dotted lines.  The blue dashed lines represent $\gamma= e^{3|g|}$. When $\gamma< e^{3|g|}$, none of the third and fourth bands and bound state \eqref{wave functionbound} exist. The red dotted-dashed lines represent $g_1= g_2$, where $g_{1}= \ln h(1/\gamma-\gamma-U)$ and $g_{2}= \ln\left[ \gamma/h(1/\gamma-\gamma-U)\right]/2$ with $h(x)=(\sqrt{x^2+4}+|x|)/2$. The black cross represents $(\gamma, U) = (2.855434,-2)$ and the black circle represents $(\gamma, U) = (5,-2)$. The three green squares in (c) represent $(\gamma, U) = (5,-3.5)$, $(5,-5.8)$, and $(7,-5.8)$, which are points in the regions $B_1$, $A$, and $B_2$, respectively. (e) The spectra of the system with parameters $(g,\gamma,U)=(0.25,5,-5.8)$ and $(g,\gamma,U)=(0.25,5,-5.8)$. }
\label{fig3}
\end{figure*}

 With wave function  $|\psi \rangle=\sum_{x_1\leqslant x_2 }\left[\sqrt{2}- (\sqrt{2}-1)\delta_{x_1,x_2}  \right] u(x_1,x_2) \hat{a}_{x_1}^{\dag}\hat{a}_{x_2}^{\dag}| 0 \rangle$, the eigenvalue equation $H|\psi \rangle =E|\psi \rangle$ can be rewritten as homogeneous linear difference equation about $u(x_1,x_2)$:
 \begin{align}\label{Hami}
-\sum_{a=\pm 1} te^{ag}\left[u(x_1+a,x_2)+u(x_1,x_2+a)\right]& \\
 +U\delta_{x_1,x_2} u(x_1,x_2)= E& u(x_1,x_2), \notag
\end{align}
and the equations of GBCs are
\begin{align}\label{GBC1}
-\sum_{a=\pm 1} te^{ag}u(0,x_2+a) +U\delta_{0,x_2} u(0,x_2)& \\
-\gamma e^g u(1,x_2)-e^g u(-1,x_2)= E& u(0,x_2), \notag
\end{align}
and
\begin{align}\label{GBC2}
-\sum_{a=\pm 1} te^{ag}u(1,x_2+a) +U\delta_{1,x_2} u(1,x_2)& \\
- e^g u(2,x_2)-\gamma e^g u(0,x_2)= E& u(1,x_2). \notag
\end{align}
In the absence of interaction ($U=0$), the model reduces to the Hatano-Nelson (HN) model \cite{Hatano96,Hatano97} with GBC
\cite{LiLH2021,CXGuo2021}, which can be exactly solvable in the whole parameter region \cite{CXGuo2021}.  The boundary parameter $\gamma$ interpolates the OBC($\gamma=0$) to PBC ($\gamma=1$).

Our goal is to address the eigenvalue problem, with a focus on the bound states where both particles are confined near the boundary sites at $x=0$ and $1$. To get a straightforward view of the BIC, we display the spectra of the system with $\gamma=5$ in Fig. \ref{fig1}(b), including four complex continuum bands \cite{SM}, a BIC and two separated bound states above or below the bands, where eigenvalues of the three bound states are all real. The spectral of the first band is given by $2\cos (k_1-i g)+2\cos (k_2-i g)$ with $k_1,~k_2 \in \left[0,2\pi \right)$ and interval of real parts in $\left[-4\cosh g,~4\cosh g\right]$, corresponding to scattering states of two particles in the HN lattice. In comparison with free particles,  the scattering eigenstates have phase shifts at boundaries and the location where the two particles interact. A typical distribution of random chosen wave function in this band is displayed in the first row of Fig. \ref{fig1}(c).
The spectral of the second band is given by $-\sqrt{U^2+16\cos^2(K-ig) }$ for $U<0$ and $\sqrt{U^2+16\cos^2(K-ig) }$ for $U>0$ with $K \in \left[0,2\pi \right)$, which corresponds to the delocalized molecule states. The interval of the real part is $\left[-\sqrt{U^2+16\cosh^2 g },~-\sqrt{U^2-16\sinh^2 g }\right]$ for $U<-4\sinh |g|$ and $\left[\sqrt{U^2-16\sinh^2 g},~\sqrt{U^2+16\cosh^2 g }\right]$ for $U>4\sinh |g|$.
The density distribution of a typical state in this band is presented in the second row of Fig. \ref{fig1}(c). It is shown that two particles are bound together and distribute along the diagonal line. Here the asymmetric distribution along the diagonal line is attributed to the nonreciprocal hopping. States in the third and fourth bands correspond to scattering states with one particle bound around the boundary, with the corresponding density distribution shown in the third row of Fig. 1(c).  The energy of the bound particle can be either of $\pm (\gamma + 1/\gamma )$, while the other is $2\cos (k-i g)$ with $k \in \left[0,2\pi \right)$. Thus the real part of these two bands covers   $\left[\gamma + 1/\gamma-2\cosh g,~\gamma + 1/\gamma+2\cosh g\right]$ and $\left[-\gamma - 1/\gamma-2\cosh g,~-\gamma - 1/\gamma+2\cosh g\right]$, respectively.

To characterize the distribution properties of states in these four bands and bound states, we can numerically calculate their fractal dimension (FD) defined as
 \begin{equation}\label{FD}
\text{FD}_j=-\ln (\text{IPR}_j)/\ln(N),
\end{equation}
where $\text{IPR}_j =\sum_{x_1,x_2}|u_j(x_1,x_2)|^4$ is the inverse participation ratio. The FDs  for the states in the first band, in the other bands, and bound states are displayed in Fig. \ref{fig1} (b).  In the limit of $ N \rightarrow\infty$, they approach to $2$, $1$ and $0$, respectively, which can be obtained by finite size analysis \cite{SM}.
To see it more clearly, we also display the corresponding IPRs with different lattice sizes in Fig. \ref{fig1}(d). It is shown that the IPRs of states in continuum bands change with the lattice size, whereas IPRs of the three bound states are not sensitive to the change of lattice size. A BIC corresponds to the bound state with energy falling in $\left[-4\cosh g,~4\cosh g\right]$.

{\it Analytical solution  of BIC.-}
Although general eigenstates of our model (1) are not solvable, we show that the BIC can be analytically derived by taking the Bethe ansatz type (BAT) wave function \textcolor{red}{\cite{SM} }
\begin{eqnarray}
u_{b}(x_1,x_2) = u_{b,0}(x_1,x_2)  e^{-g(x_1+x_2)},
\label{BICWF}
\end{eqnarray}
where
\begin{eqnarray}
u_{b,0}(x_1,x_2) =
\left\{
\begin{array}{cc} \label{LEG}
f_1(x_1,x_2) ,~~~~~~& 0<x_1\leqslant x_2  \\
f_2(x_1,x_2), ~~~~~~&x_1\leqslant0 < x_2,\\
-f_1(-x_2,-x_1) , & x_1\leqslant x_2 \leqslant 0 \\
f_1(x_2,x_1) ,~~~~~~& 0<x_2<x_1,\\
f_2(x_2,x_1),~~ ~~~~&x_2\leqslant0 < x_1,\\
-f_1(-x_1,-x_2) ,& x_2<x_1 \leqslant 0, \\
\end{array} %
\right.
\end{eqnarray}
with
\begin{align}\label{coef3}
f_1&(x_1,x_2) = \frac{ e^{ik_{1}}-\gamma^2  e^{-ik_{1}}}{2i\gamma \sin k_{1}} e^{ik_1 x_1 +ik_2 x_2} \\
 &+\frac{ e^{ik_{1}}(\gamma^2 -1)}{2i\gamma \sin k_{1}}e^{-ik_1 x_1 +ik_2 x_2}-\frac{ e^{ik_{1}}(1-\gamma^2)}{2i\gamma \sin k_{2}}e^{ik_2 x_1 -ik_1 x_2}, \notag
\end{align}
and
 \begin{equation}\label{wave functionbound}
f_2(x_1,x_2) = e^{ik_1 x_1 +ik_2 x_2}-e^{ik_1+ik_2}e^{-ik_2 x_1 -ik_1 x_2} .
\end{equation}
\textcolor{red}{}

To correctly describe a bound state, from \eqref{BICWF}, we see that $|e^{ik_1}|>e^{|g|}$ and $|e^{ik_2}|<e^{-|g|}$ are required. Meanwhile, $k_{1,2}$ are given by
\begin{align}\label{k12}
\gamma= e^{-ik_{2}},~~~1/\gamma-\gamma-U=e^{ik_{1}}-e^{-ik_{1}}.
\end{align}
The exact energy of bound state $E_{b}=-2\cos k_1 -2\cos k_2$ has two possibilities: $E_{b1}=-1/\gamma-\gamma-\sqrt{(\gamma-1/\gamma+U)^2+4}$ for $1/\gamma-\gamma>U$ and $E_{b2}=-1/\gamma-\gamma+\sqrt{(\gamma-1/\gamma+U)^2+4}$ for $1/\gamma-\gamma<U$. Here, we can see that the energy of the bound state is independent of the parameter $g$. However, the wave function is different from the Hermitian case by a factor of $e^{-g(x_1+x_2)}$.

{\it Phase diagram of BIC.-}
To maintain the state of $E_{b}=E_{b1}$ as a bound state, the condition $U_{1,1}<U<U_{1,2}$ must be satisfied, where $U_{1,1}=2(e^{|g|}/\gamma-e^{-|g|}\gamma)\cosh g$ and $U_{1,2}=1/\gamma-\gamma-2\sinh |g|$, which forms areas $A$. This region shrinks with increasing $g$, as shown in Figs. \ref{fig3} (a)-(d). Since $E_{b1}<\min\{-4\cosh g,~-\sqrt{U^2+16\cosh^2 g },~\pm(\gamma + 1/\gamma)-2-2\cosh g\}$, the bound state is below all four continuum bands in the real axis and thus is not a BIC. A typical spectrum structure for the system marked by the square in the region of $A$ is displayed in Fig.  \ref{fig3} (e1). The two transfer points $U_{1,1}$ and $U_{1,2}$ are determined by the touching points that $E_{b1}$ touches the second and fourth continuum bands, respectively.
 \begin{figure}[tbp]
\includegraphics[width=0.45\textwidth]{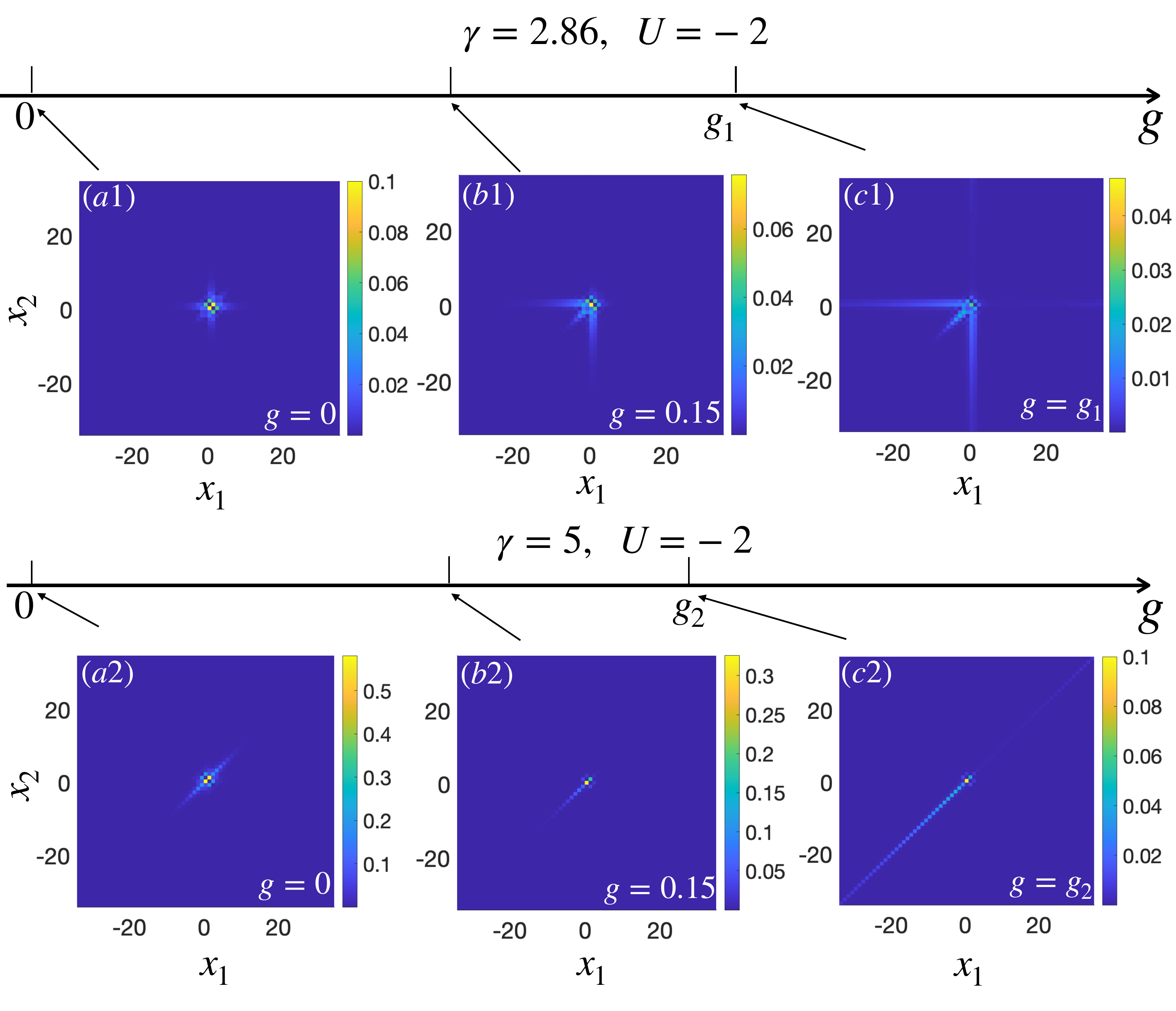}
 \caption{(a1)-(c1) The density distribution of the BIC with $g=0$, $g=0.15$, and $g=g_1$  for the system with $\gamma=2.86$ and $U=-2$. (a2)-(c2) The density distribution of the BIC with $g=0$, $g=0.15$, and $g=g_2$ for the system with $\gamma=5$ and $U=-2$.}
\label{dis-BIC}
\end{figure}

Similarly, the condition to maintain the state of $E_{b}=E_{b2}$ as a bound state is $U_{2,1}<U<U_{2,2}$ with $U_{2,1}=1/\gamma-\gamma+2\sinh |g|$ and $U_{2,2}=-2(e^{|g|}/\gamma+e^{-|g|}\gamma)\sinh|g|$, which forms areas $B_{1,2}$ as shown in Fig. \ref{fig3}. As the value of $g$ increases, the corresponding regions also shrink. The points $U=U_{2,1}$ and $U=U_{2,2}$ are just the touching points of $E_{b2}$ with second and fourth bands. We can prove that $E_{b2}>\max\{-\sqrt{U^2+16\cos^2(\pi/2 -ig) }~,-(\gamma+1/\gamma)+2\cosh (g) \}$ and $E_{b2}<(\gamma+1/\gamma)-2\cosh (g) $. Thus this bound state cannot fall in the second to fourth bands. It may fall within the first continuum band, if an additional condition $E_{b2}>-4\cosh g$ is fulfilled, which gives rise to the area $B_1$ as displayed in Fig. \ref{fig3}. In the parameter region $B_1$  there exists a BIC with a typical spectrum structure depicted in Fig. \ref{fig1}(b). As a comparison, a typical spectrum for the system marked by the square in region of $B_2$ is displayed in Fig. \ref{fig3} (e2).

Although the parameter $g$ does not change the energy of the bound state, it changes the continuous spectrum and affects the fate of BIC by modifying the wave function. With the increase in $g$, the first band expands in both directions along the real axis, while the second and fourth bands also increase in size.  When the $2nd$ or $4th$ band touches the bound energy, the bound state merges into this band and vanishes. The touching point gives rise to a threshold value $g_{1}= \ln h(1/\gamma-\gamma-U)$ with $h(x)=(\sqrt{x^2+4}+|x|)/2$ or $g_{2}= \ln\left[ \gamma/h(1/\gamma-\gamma-U)\right]/2$, determined by either $E_{b2}=-\gamma-1/\gamma+2\cosh |g_1|$ or $E_{b2}=-\sqrt{U^2-16\sinh^2 g}_2$.
In the parameter space spanned by $\gamma$ and U, the relation $g_1(\gamma,U)=g_2(\gamma,U)$ gives a dividing line distinguishing which band expanding to $E_{b2}$ first,  as shown in Figs. \ref{fig3} (a)-(d) by the dashed-dotted red lines.
The bound state above the dividing line merges into the fourth band and vanishes when $g>g_1$, whereas the bound state below the dividing line merges into the second band when $g>g_2$. Besides the threshold values determined by different relations, the BICs above and below the dividing line exhibit different spatial distributions, as displayed in Figs. \ref{dis-BIC} (a1)-(c1) and Figs. \ref{dis-BIC} (a2)-(c2), respectively. Here, density distributions shown in Figs. \ref{dis-BIC} (a1)-(c1) correspond to the circles in Figs. \ref{fig3} (a)-(c), and the ones in Figs. \ref{dis-BIC} (a2) and (b2) correspond to the crosses in Figs. \ref{fig3} (a) and (b), respectively.


With the increase in $g$, it can be found from Fig. \ref{dis-BIC} that the BIC is delocalized either along the $x_1$ and $x_2$ axis or the diagonal direction $x_1=x_2$. The wave function along the axis-$x_1$ is $u_{b}(x_1,0) =e^{-gx_1}u_{b,0}(x_1,0)$ with
\begin{eqnarray}
u_{b,0}(x_1,0) =
\left\{
\begin{array}{cc} \label{wavex10}
e^{ik_2x_1}-e^{ik_1+ik_2}e^{-ik_1x_1}   ,& x_1> 0,  \\
e^{ik_1x_1 }-e^{ik_1+ik_2}e^{-ik_2 x_1 }, & x_1\leqslant 0.\\
\end{array} %
\right.
\end{eqnarray}
Equation \eqref{wavex10} is the superposition of two exponentially decaying wave functions $e^{-ik_1x_1}$ and $e^{ik_2x_1}$ with exponential decay factors (EDFs)  $L_{h1}=-\text{Im}(k_1)$ and $L_{h2}=\text{Im}(k_2)$, respectively. From Eq. \eqref{k12}, it follows that $L_{h1}= \ln h(1/\gamma-\gamma-U)$ with $h(x)=(\sqrt{x^2+4}+|x|)/2$ \cite{ik1} and $L_{h2}=\ln \gamma$.
Since $L_{h1}<L_{h2}$ is always fulfilled for BIC, we see that $u_{b}(x_1,0)$ is completely delocalized when $g=L_{h1} \equiv g_1$.  The wave function $u_{b}(0,x_2)$ has the same properties due to the exchange symmetry of $x_1$ and $x_2$.
The wave function \eqref{LEG} along the diagonal line $x_1=x_2$, i.e., $u_{b,0}(x_1,x_1)$, also has two EDFs: $L_{d1}= \text{Im}(k_1+k_2)= \ln\left[ \gamma/h(1/\gamma-\gamma-U)\right]$ and $L_{d2}=\text{Im}(k_2 -k_1)=\ln \left[ \gamma h(1/\gamma-\gamma-U)\right]$. Since $L_{d1}<L_{d2}$,  $u_{b}(x_1,x_2)$ is completely delocalized along the diagonal line when $g=L_{d1}/2 \equiv g_2$. From the analysis of wavefuntion of BIC, we see clearly that the threshold value for the breakdown of BIC is given by  $g=\min\{g1,g2\}$.

\begin{figure}[tbp]
\includegraphics[width=0.45\textwidth]{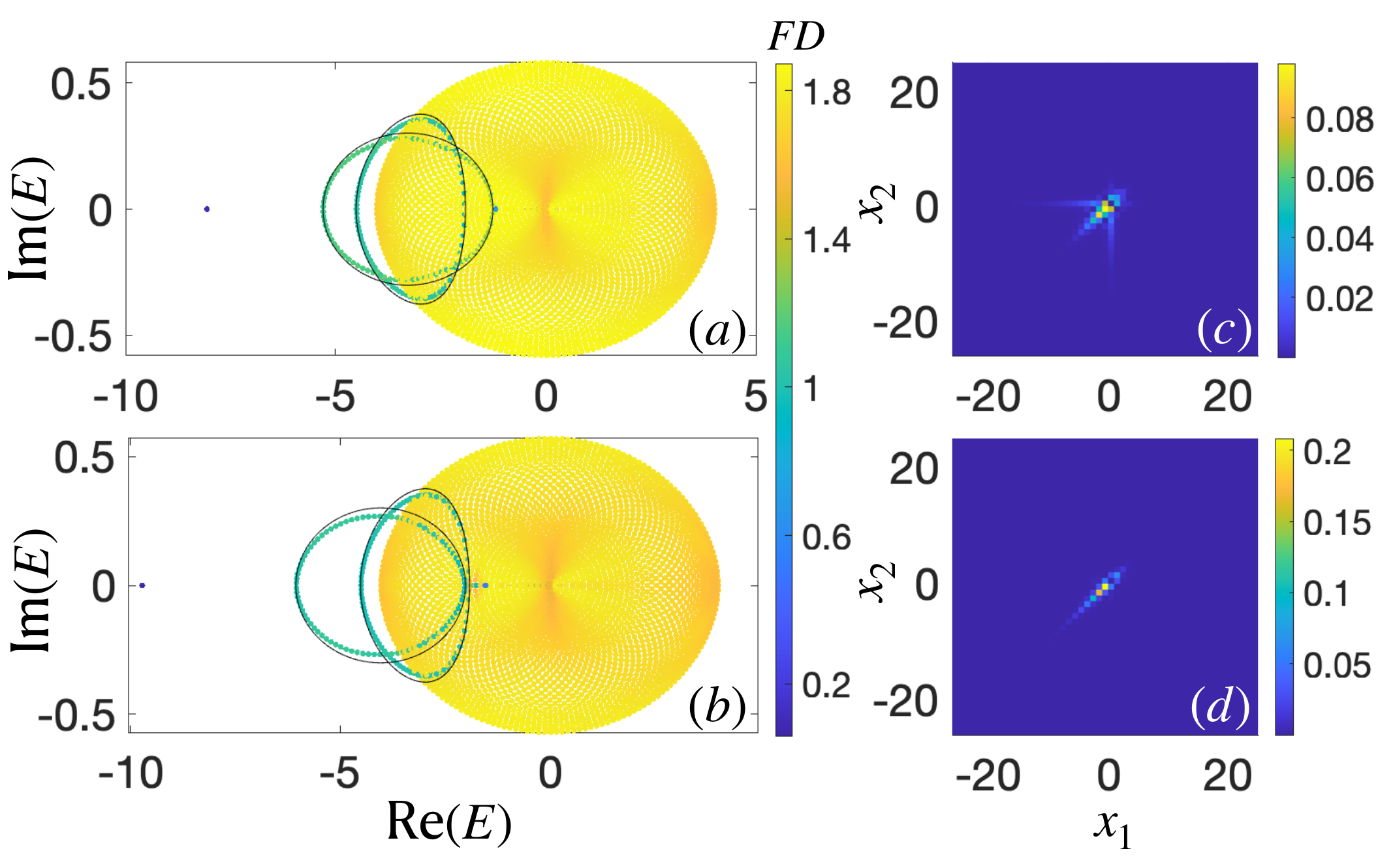}
 \caption{(a) (b) The real and imaginary parts of spectra of the system with parameters $(g,V,U)=(0.15,-2.6,-2)$ and $(g,V,U)=(0.15,-3.5,-2)$, respectively. The color of each dot represents the fractal dimension FD of different eigenstates. The two black circles from left to right are the third band $E_{3}=-\sqrt{V^2+4}+2\cos (k-i g)$ and the second band $E_{2}=-\sqrt{U^2+16\cos^2(K-ig) }$, respectively. (c) (d) The density distribution of the BIC for the system with parameters $(g,V,U)=(0.15,-2.5,-2)$ and $(g,V,U)=(0.15,-3.5,-2)$, respectively. The length of the lattice is $2M+1=91$.}
\label{fig4}
\end{figure}

{\it BIC in the impurity model.-}
Next we show that BIC can be also identified in the two-particle Hatano-Nelson model with an impurity under periodic boundary condition, described by
 \begin{align}\label{Hamiltonian2}
 \hat{H}=&\sum_{x=-L} ^{M}  \left [ -e^{g} \hat{a}_x^{\dag}  \hat{a}_{x+1} - e^{-g} \hat{a}_{x+1} ^{\dag}  \hat{a}_x  +\frac{U}{2} \hat{a}_x^{\dag}  \hat{a}_x^{\dag}  \hat{a}_x  \hat{a}_x \right] \notag\\
& +V \hat{a}_0^{\dag}  \hat{a}_{0} ,
 \end{align}
 where the impurity strength $V<0$ and $U<0$. Similarly, BIC can be solved exactly by applying the BA method. The corresponding energy $E_{BIC}=-\sqrt{V^2+4}+\sqrt{(V-U)^2+4}$ and the conditions for existence of BIC are $V+2\sinh |g|<U< V-e^{-2|g|}V-2\sinh^2(2|g|)\frac{V+\sqrt{V^2+4}}{2}$ and $E_{BIC}>-4\cosh g$. For this system, there are three continuum bands, the first band $2\cos (k_1-i g)+2\cos (k_2-i g)$ with $k_1,~k_2 \in \left[0,2\pi \right)$, the second band $\text{sign}(U)\sqrt{U^2+16\cos^2(K-ig) }$ with $K \in \left[0,2\pi \right)$, and the third band $-\sqrt{V^2+4}+2\cos (k-i g)$, with $k \in \left[0,2\pi \right)$, as shown in Figs. \ref{fig4} (a) and (b).

 The BICs also exhibit two kinds of different distributions, as displayed in  Figs. \ref{fig4} (c) and (d), leading to different ways to compete with the nonreciprocal term. The first way is that the increase of $g$ causes the BIC to lose localization along the axis of $x_1$ and $x_2$ when $g_{1}(V,U)<g_{2}(V,U)$, where $g_{1}(V,U)=\ln h(V-U)$ and $2g_{2}(V,U)=\ln[h(V)/h(V-U)]$ are the minimum EDFs of the wave function of BIC along the $x_1$ and $x_2$ axis and the diagonal direction $x_1=x_2$ with $g=0$. This BIC will merge into the third band. The second way is that as the value of $g$ increases, the wave function of BIC becomes extended along the diagonal direction when $g_{2}(V,U)<g_{1}(V,U)$. This BIC will merge into the second band.


{\it Summary and discussion.-}
Using the Bethe ansatz method, we have obtained the exact solution for the BIC in the two-particle interacting Hatano-Nelson model with either generalized boundary conditions or an impurity potential. Our results demonstrate that the interplay of interactions, boundary potential and nonreciprocal hopping can give rise to two types of BICs with different spatial distributions. The exact wave function and energy of BIC enable us to get a precise phase diagram of BICs with the boundaries marking the emergence and breakdown of BICs being analytically determined.
In principle, the BIC may exist in many-particle interacting systems with impurity, although no analytical results are available. Numerically, we demonstrate the existence of BIC in three-particle system, which can survive in some parameter regions even in the presence of three-particle interaction \cite{SM}.
\\

\begin{acknowledgments}
We would like to thank Y.-S. Cao for useful discussions.
Y Liu is supported by the National Natural Science Foundation of China Grant No. 12204406.
S Chen is supported by National Key Research and Development Program of China (Grant No. 2023YFA1406704 and 2021YFA1402104), the NSFC under Grants No. 12174436 and No. T2121001, and the Strategic Priority Research Program of the Chinese Academy of Sciences under Grant No. XDB33000000.
\end{acknowledgments}
\appendix

%

\begin{thebibliography}{36}
\bibitem{ik1} Let $e^{ik_{1}}=\beta$, Eq. \eqref{k12} becomes $1/\gamma-\gamma-U=\beta-1/\beta \rightarrow  0=\beta^2- (1/\gamma-\gamma-U)\beta -1$. The formula of roots of a quadric equation gives $\beta_{\pm}=\frac{1/\gamma-\gamma-U\pm \sqrt{(1/\gamma-\gamma-U)^2+4}}{2}$. Because $1/\gamma-\gamma<U$ and $|e^{ik_1}|>e^{|g|}$, we only can choose
$\beta = \beta_{-}=\frac{1/\gamma-\gamma-U- \sqrt{(1/\gamma-\gamma-U)^2+4}}{2}$. Then we can get that $i k_1= \ln \beta_{-} = \ln h(1/\gamma-\gamma-U)+i\pi$ with $h(x)=(\sqrt{x^2+4}+|x|)/2$.
\end{thebibliography}

\end{document}